\documentclass[%
reprint,
superscriptaddress,
amsmath,amssymb,
]{revtex4-1}

\usepackage{graphicx}
\usepackage{subfigure}
\usepackage{dcolumn}
\usepackage{bm}
\usepackage{longtable}
\usepackage{threeparttablex}
\usepackage{amssymb}
\usepackage{amsthm}
\usepackage{amsmath}
\usepackage{lineno}
\usepackage{hyperref}

\begin{document}
	
\preprint{arXiv.org}
	
\title{Development of a continuously tunable titanium-sapphire laser system for the ARIEL laser ion source}
	
\author{Bianca Bettina Reich}
\email{B.B.Reich@gsi.de}
\affiliation{TRIUMF-Canada's Particle Accelerator Center, Vancouver, BC V6T 2A3, Canada}
\affiliation{Department of Physics and Astronomy, University of Heidelberg, 69120 Heidelberg, Germany} 
        
\author{Maryam Mostamand}
\affiliation{TRIUMF-Canada's Particle Accelerator Center, Vancouver, BC V6T 2A3, Canada}
\affiliation{Department of Physics and Astronomy, University of Manitoba, Winnipeg MB, Canada, R3T 2N2} 
            
\author{Ruohong Li}
\affiliation{TRIUMF-Canada's Particle Accelerator Center, Vancouver, BC V6T 2A3, Canada}

\author{Jens Lassen}
\affiliation{TRIUMF-Canada's Particle Accelerator Center, Vancouver, BC V6T 2A3, Canada}
\affiliation{Department of Physics and Astronomy, University of Manitoba, Winnipeg MB, Canada, R3T 2N2}
\affiliation{Department of Physics, Simon Fraser University, Burnaby, BC, Canada, V5A 1S6}

	
\begin{abstract}
	A concept for continuously tunable titanium-sapphire (Ti:Sa) lasers using dispersion prisms is under investigation for the ARIEL (Advanced Rare IsotopE Laboratory) laser ion source at TRIUMF (Canada's particle accelerator center). Wavelength selection for pulsed Ti:Sa lasers used in hot cavity laser resonance ionization spectroscopy is usually done with birefringent filters (BRFs) and etalons or diffraction gratings. For resonance ionization spectroscopy a laser system allowing a continuous wavelength scan is necessary. Tunable lasers based on BRFs and etalons have high output powers however require synchronized optimization for continuous laser wavelength scans and are therefore laborious to use in scanning applications. Diffraction grating tuned lasers can provide continuous wavelength scan over 200~nm range but typically have lower output laser power due to the grating deformation under high pumping power. Aiming to overcome both shortcomings a laser design based on prisms as dispersing element has been revisited. Simulations on the beam path and optical reflectivity are done which show that these losses can be minimized to around 0.04~\% for a tuning range from 700~nm up to 920~nm. Further improvement on the tuning range and reduction on the linewidth will be pursued.  
		
	\keywords{titanium-sapphire laser \and tunability \and prisms}	

\end{abstract}
	
\maketitle

\section{Introduction}\label{intro}
In the present paper an improved tuning method of the Ti:Sa lasers for the Resonant Ionization Laser Ion Source (RILIS) at TRIUMF is being reported. For on-line mass separator facilities the selective production of radioactive ion beams (RIB) is most successfully done by resonant laser excitation and ionization \cite{Jens2005}. A huge advantage of a RILIS is that the laser system is accessible during RIB production and that it can be transferred to more than one target station. The new Advanced Rare IsotopE Laboratory (ARIEL) currently under construction at TRIUMF is a multidisciplinary facility ranging from nuclear science over industrial manufacturing to medical science. It produces rare isotopes which normally only occur in stars while they burn or explode. ARIEL plans to deliver up to three simultaneous RIB. Therefore a robust and easily tunable laser system with high output powers is desired.\\

The wavelength selection for pulsed Ti:Sa lasers is usually done with BRF or diffraction gratings. BRF-tuned lasers normally have a higher output power due to lower cavity loss compared to the lasers with diffraction gratings but make a continuous wavelength scan beyond 0.2~nm laborious \cite{Ruohong2017continuouslytuneable}. To achieve a continuous wavelength scan, a grating-tuned laser was designed and built with gold coated 500~nm-blazed grating \cite{Teigelhofer2010}. This setup has two shortcomings: first reflection losses occur due to the gold coating and second the intra-cavity power and therefore the overall output power is limited by local heating and deformation of the grating \cite{Ruohong2017continuouslytuneable}. A dual-cavity design has been introduced to avoid these grating deformation issues boosting up the laser power to 1.5~-~2 times high. Nevertheless grating-tuned lasers still underperform BRF-tuned lasers in aspects of laser power and stability especially at the two ends of the Ti:Sa crystal emission curve under 720~nm and over 900~nm, if not changed to specially coated cavity mirrors.\\

This work aims to develop an alternate laser system with dispersion prisms similar to \cite{Spie2013,Jungbluth2010} for both wide and easy tunability and robust operation with high output powers in order to meet the needs of the coming ARIEL era.

\section{General setup}\label{sec:setup}
First experiments are done with a setup of five prisms. As it will be shown the reflectivity losses increase with each prism which is why a second round of experiments is done with only four prisms. To get the same total deviation of the beam prisms with a higher refractive index are used. The alignment of the prisms is optimized for a wavelength of 800~nm.\\

The Nd:YAG pump laser (\textit{Nanio 532} from \textit{InnoLas Photonics}) has a nominal power of 18 W at 10 kHz repetition rate and a quality factor of M$^{2}$ $<$ 1.3. The beam is forwarded by two coated dielectric high reflection (HR) mirrors and focused by a plano-convex lens with a focal length of 75~mm before entering the cavity. The Ti:Sa crystal has an absorption coefficient of 1.60 and is situated inside the cavity between two curved mirrors with broadband dielectric coatings with HR $>$ 97.7~\% for 650~nm to 1040~nm and anti-reflection (AR) $<$ 0.2~\% for 532~nm. The crystal is placed off-center cavity towards the first curved mirror so that the pump focus is situated right behind the crystal. The broadband output coupler has a reflectivity of 80~\% for a wavelength range from 650~nm to 1100~nm and is situated behind the second curved mirror. After the first curved mirror five (four) prisms followed by a high reflection mirror are placed. While the beam passes the prisms dispersion occurs leading to a separation of different wavelengths. At the end of the prisms a rotatable high reflection mirror is placed to select a wavelength and reflect it back into the cavity. The optical elements are all made from fused silica and have a surface flatness of $\lambda$/10 at 633~nm standard.

\begin{figure}
	\centering
	\includegraphics[width=0.49\textwidth]{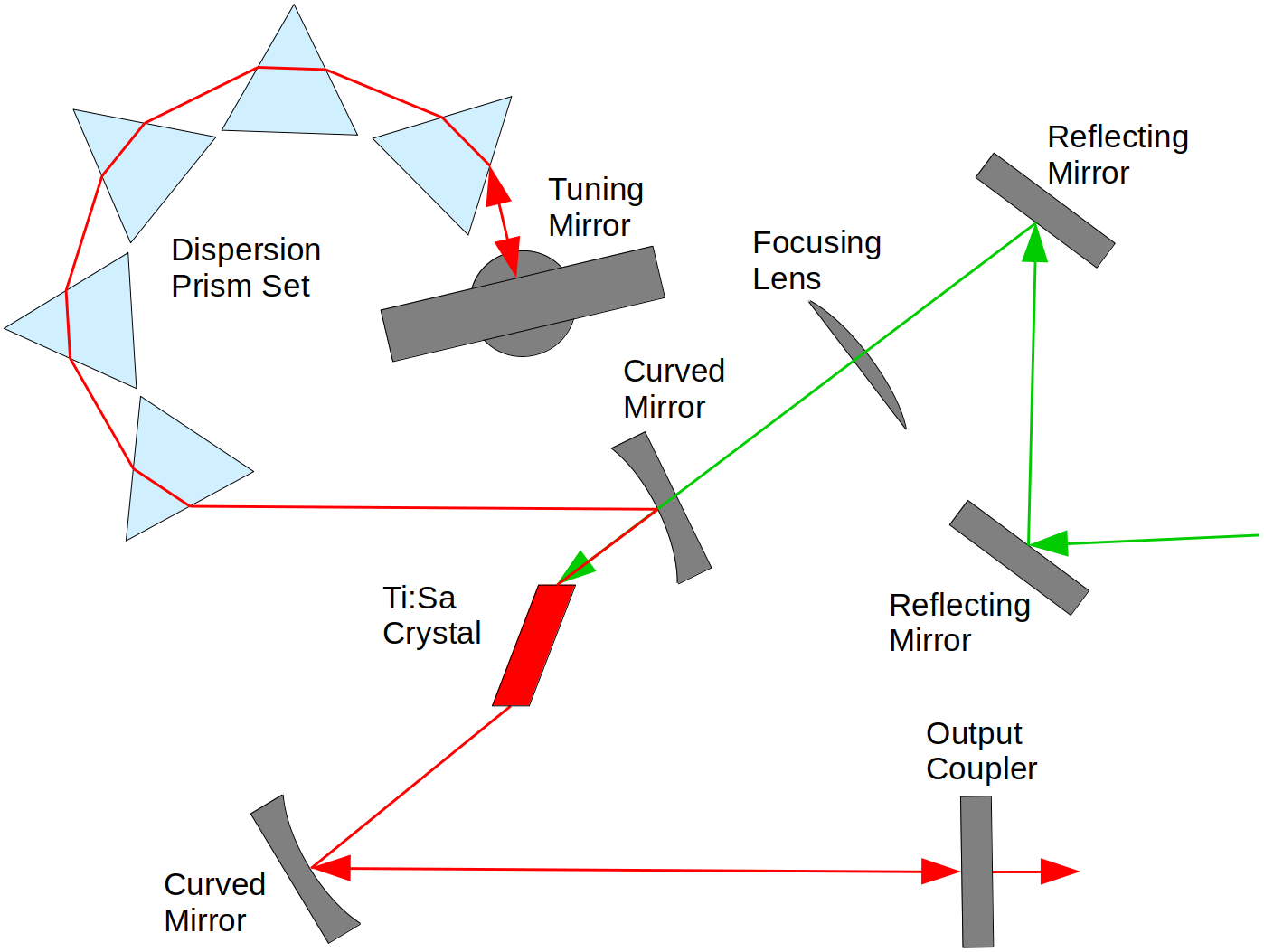}
	\caption{Schematic diagram of a mirror-tuned Ti:Sa laser using prisms as dispersive elements. The Nd:YAG laser beam (green) hits the Ti:Sa crystal (red) whereby it starts lasing. The laser beam of the Ti:Sa laser (red) is deflected by five prisms creating dispersion. The rotatable tuning mirror selects the wanted wavelength which gets reflected back into the cavity.}
	\label{NewSetup}      
\end{figure}

\section{Mode matching}

\begin{figure}
	\centering
	\includegraphics[width=0.49\textwidth]{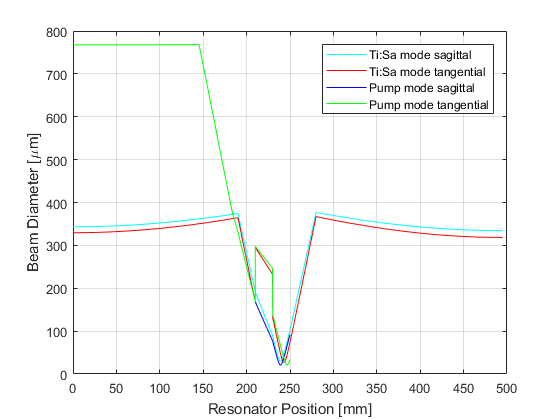}
	\caption{Mode matching with a thin lens.
		One can see the first curved mirror at around 190~mm inside the resonator. The tangential plane for the pump mode and the Ti:Sa mode show the crystal at around 210~mm to 230~mm. Roughly at 240~mm one can see that the beam diameter reaches its smallest value which is the focal point of the thin lens. At 270~mm the beam hits the second curved mirror.}
	\label{ModeMatchingLens75}
\end{figure}

Mode matching means to overlap the modes from the pump laser and the operating laser. This is essential for the operation of the laser and important for maximizing the output power. Therefore the beam path \cite{renk2012basics} and the corresponding beam diameter \cite{hodgson2005beampropagation} have to be calculated. The cavity parameters are based on \cite{Rothe2012} and are slightly modified to suit the new requirements. All calculations can be seen in \cite{Bianca2019}. During the experiment it can be seen that the Ti:Sa crystal is working as a thermal lens probably due to significant radial temperature gradients created by the pump laser which has a M$^{2}$ $<$ 1.3 and a pulse duration of 40 ns. The mode matching shown in figure \ref{ModeMatchingLens75} does not account for the effect of thermal lensing. Hence the mode matching needs to be optimized by increasing the distance between the lens and the first curved mirror. That is done in two steps first to 55~mm and second to 70~mm leading to an improved wavelength tuning range of the system.

\section{Arrangement of the prisms}

To minimize losses through reflectivity it is essential to find an optimal arrangement for the prism setup. For a transition from one material to another the angle of incidence for which light will be perfectly transmitted without reflection losses is the Brewster’s angle \cite{Demtroeder2017Experimentalphysik2}. However the Brewster’s angle is not favored when passing through a symmetric prism since the angle exiting the prism will be different. Therefore to minimize the reflectivity losses a path parallel to the prisms base is preferred \cite{Meschede2008}. This ideal angle of incidence $\alpha$ depends on the aperture angle of the prism $\gamma$ and is given by\\

\begin{equation}\label{IncidenceAngle}
\alpha = \arcsin\left(n \cdot \sin\left(\frac{\gamma}{2}\right)\right)
\end{equation}\\

where $n$ is the refractive index of the prism \cite{Meschede2008}. Through the refractive index the angle of incidence indirectly depends on the wavelength. Therefore the angles of incidence and emergence after every prism will differ for all wavelengths for which the setup was not optimized for and need to be calculated iteratively shown in \cite{Bianca2019}. The level of reflectivity of light hitting a surface can then be determined by Fresnel’s equations depending on its polarization \cite{nolting2011elektrodynamik}. For p-polarized light the reflectivity losses $R_{p}$ are given by\\

\begin{equation}\label{ReflectivityFresnel}
R_{p} = \left(\frac{\left(\frac{n_{2}}{n_{1}}\right)^{2} \cdot \cos(\alpha) - \sqrt{\left(\frac{n_{2}}{n_{1}}\right)^{2} - \sin^{2}(\alpha)}}{\left(\frac{n_{2}}{n_{1}}\right)^{2} \cdot \cos(\alpha) + \sqrt{\left(\frac{n_{2}}{n_{1}}\right)^{2} - \sin^{2}(\alpha)}}\right)^{2}
\end{equation}\\

where $n_{1}$ is the refractive index of air and $n_{2}$ is the refractive index of the prism \cite{nolting2011elektrodynamik}. Additionally the refractive index of the prisms is of great interest for the optimal arrangement. The first experiments are done with five prisms made from fused silica. Their refractive index is 1.45 at a wavelengths of 800~nm \cite{Malitson:65} for which the setup is optimized. A second round of experiments is done with four prisms made out of SF10 (dense flint) and a refractive index of 1.71 at a wavelength of 800~nm \cite{SF10}. The optimal angle of incidence is calculated with equation (\ref{IncidenceAngle}). Therefore the angle between the prisms is 41.74$^\circ$ for the five prism setup and 58.79$^\circ$ for the four prism setup.

\section{Reflectivity}

\begin{figure}[!h]
	\subfigure[Reflectivity of five prism setup]{\includegraphics[width=0.49\textwidth]{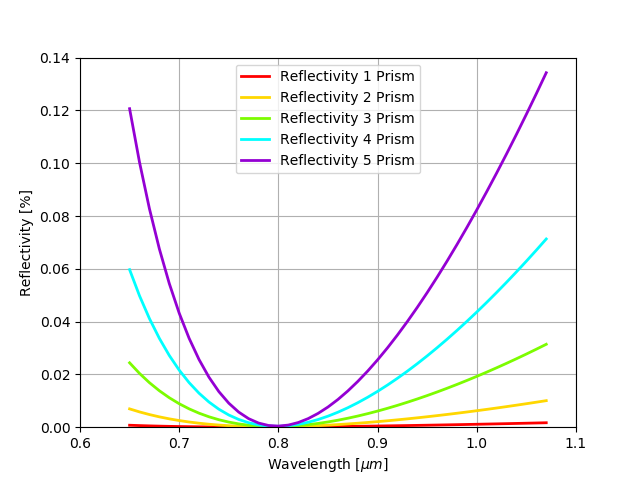}}
	\subfigure[Reflectivity of four prism setup]{\includegraphics[width=0.49\textwidth]{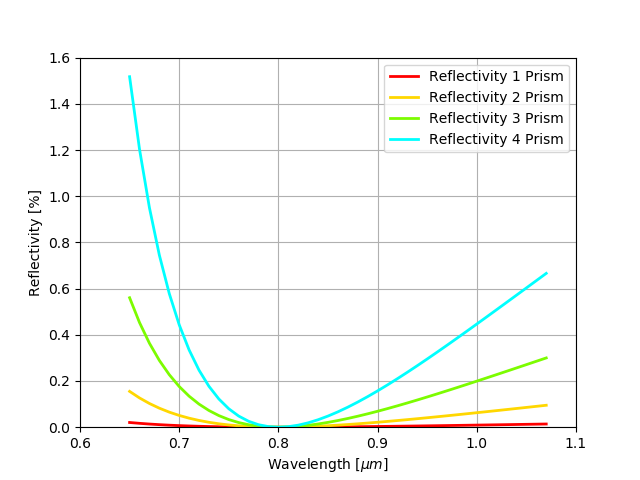}}
	\caption{It can be seen that the system is optimized for a wavelength of 800~nm and that the reflectivity increases with each prism. In figure (a) the highest reflectivity is reached for a wavelength of 1070~nm and is around 0.13~\%. In figure (b) the highest reflectivity losses are reached for a wavelength of 650~nm and is around 1.5\%. It can be seen that due to the higher refractive index of the prisms the reflectivity losses increase in respect to the five prism setup even though the setup has one prism less.}
	\label{ReflectivityFigure}
\end{figure}

The losses through reflectivity are calculated with equation (\ref{ReflectivityFresnel}) and depend on the changing angle of incidence. This angle is different for every wavelength and differs more and more with every prism passed. The reflectivity losses shown in figure \ref{ReflectivityFigure} cover the theoretical tuning range of a Ti:Sa laser from 650~nm to 1070~nm. The current setup is tunable from around 700~nm to 920~nm and therefore has much lower reflectivity losses. The five prism setup is shown in figure \ref{ReflectivityFigure}(a) and the four prism setup in figure \ref{ReflectivityFigure}(b). In the case of the five prism setup the highest reflectivity losses are 0.04~\% for a wavelength of 920~nm. For the four prism setup the highest reflectivity losses are 0.3~\% for a wavelength of 730~nm.

\section{Tuning range}

\begin{figure}
	\subfigure[Tuning range of five prism setup]{\includegraphics[width=0.49\textwidth]{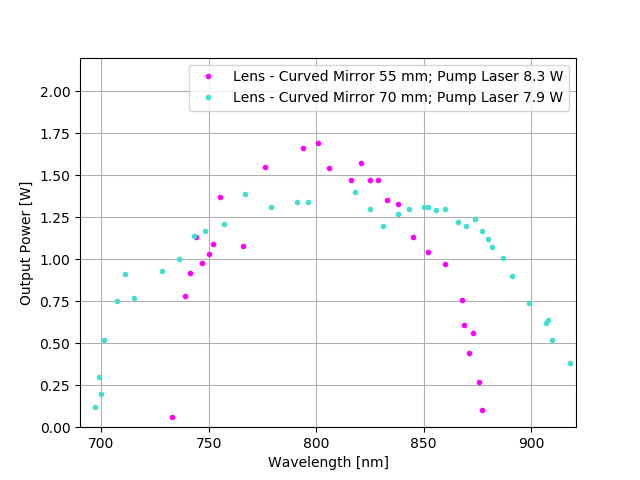}}
	\subfigure[Tuning range of four prism setup]{\includegraphics[width=0.49\textwidth]{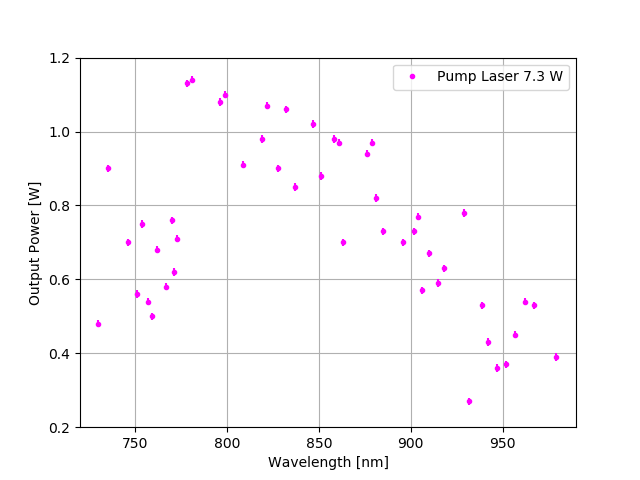}}
	\caption{The maximum output power for both configurations is accomplished at 800~nm. In figure (a) it can be seen that the tuning range increases with an increasing distance between the lens and the first curved mirror. At a distance of 55~mm the tuning range is roughly 150~nm wide. At a distance of 70~mm the tuning range increased to around 220~nm. In figure (b) the tuning of the laser system is very unstable. Nevertheless it shows a tuning range of roughly 240~nm.}
	\label{TuningRangeFigure}
\end{figure}

Since the pump laser is operating unstable in the region between 7.5~W and 10~W the pumping power was adjusted for every experiment. Therefore it is increased as long as the output power at a wavelength of 800~nm increases too. Additionally an etalon is placed between the second curved mirror and the output coupler. With a subsequently optimized mode matching the tuning range is increased to around 220~nm reaching from 700~nm to 920~nm which is shown in figure \ref{TuningRangeFigure}(a). Furthermore one can see that the gain of the system at a wavelength of 800~nm is roughly 20~\% in the configuration where the lens is placed 55~mm away from the curved mirror. The linewidth in this setup is $(8 \pm 2)$~GHz for a wavelength of 800~nm. The gain for the second configuration with a distance of 70~mm between the two optical elements is around 18~\% with a linewidth of $(12 \pm 2)$~GHz; both parameters are measured at a wavelength of 800~nm.\\ 

For the setup with four prisms the tuning range is a bit higher with approximately 240~nm reaching from 730~nm to 970~nm. However the operation is not as smooth and a second etalon has to be placed inside the cavity. The tuning range is shown in figure \ref{TuningRangeFigure}(b). The gain in this setup is approximately 15~\% and the linewidth is $(14 \pm 2)$~GHz; both parameters are measured at a wavelength of 800~nm.

\section{Conclusion}

The investigated laser system seems to be promising but the implementation needs improvement. A better way of fixing the prisms on their mount needs to be found. In the presented experiments the prisms only sat on a piece of paper which made them assailable to even very small vibrations. The entire system could not run in a stable mode and therefore lost lasing very easily. The pumping source needs to be switched or the pump laser should be run at higher power and reduced afterwards with suitable optical elements to overcome the region where it operates unstable.\\

The tuning range of up to 240~nm is already very broad and could probably benefit from a more stable running system as well. To reduce the linewidth more dispersion or deviation of the beam through the prisms would be preferred. To not increase the reflection losses one should keep on working with the prisms with a lower refractive index and increase the distances between them. This would automatically mean to rearrange the optical elements in the cavity. This could additionally bring some benefits: one should focus on using the parameters given in \cite{Rothe2012} for best stability, swap the high reflector and output coupler back into their earlier positions so that the pumping beam and the laser beam emerge from opposite sides and one could fit a more suitable rotatable high reflector into the cavity. Due to space problems a mirror mount for the high reflector had to be used where the mirror sits in an arm shaped mount so that the entire arm moved back and forth and not only the angle of the mirror changed.\\

The reached gain of up to 20~\% is not bad but can certainly be improved. Due to the fact that the prisms were not fixed their positions were always slightly off the calculated ones probably leading to higher reflection losses. Furthermore when the deviation of the beam is increased by more space between the prisms one etalon or even both could be removed from the cavity which definitely would increase the gain by at least 2~\%.\\

The overall result is that the adapted laser system looks very promising but more time is required to do more measurements.

\end{document}